\newcommand*{\ATLASLATEXPATH}{latex/}
\documentclass[UKenglish,PAPER,cernpreprint,paper=a4,maketitle=true,texlive=2011]{latex/atlasdoc}

\usepackage[subfigure]{latex/atlaspackage}
\usepackage{latex/atlasbiblatex}
\usepackage{latex/atlascontribute}
\newcommand*{\met}{\ensuremath{E_{\text{T}}^{\text{miss}}}\xspace}
\newcommand*{\pt}{\ensuremath{p_{\text{T}}}\xspace}
\newcommand*{\pT}{\ensuremath{p_{\text{T}}}\xspace}
\newcommand*{\TeV}{\ifmmode {\mathrm{\ Te\kern -0.1em V}}\else
\textrm{Te\kern -0.1em V}\fi}%
\newcommand*{\GeV}{\ifmmode {\mathrm{\ Ge\kern -0.1em V}}\else
\textrm{Ge\kern -0.1em V}\fi}%
\newcommand*{\HT}{\ensuremath{H_{\text{T}}}\xspace}

\newcommand*{\gluino}{\ensuremath{\tilde{g}}\xspace}

\graphicspath{{logos/}}
\usepackage[super]{nth}
\usepackage{textcomp}
\usepackage{xspace}
\usepackage{units}
\usepackage{siunitx}
\usepackage{multirow}

\usepackage{arydshln}

\newcommand{\CLs}{CL\textsubscript{s}\xspace}

\newcommand{\mt}{\ensuremath{m_\mathrm{T}}\xspace}

\def\lsim{\mathrel{\rlap{\lower4pt\hbox{\hskip1pt$\sim$}}
    \raise1pt\hbox{$<$}}}
\def\gsim{\mathrel{\rlap{\lower4pt\hbox{\hskip1pt$\sim$}}
    \raise1pt\hbox{$>$}}}

\providecommand{\met}{\ensuremath{E_\mathrm{T}^{\mathrm{miss}}}\xspace}

\newcommand{\wjets}{$W$+jets\xspace}

\newcommand{\zjets}{$Z$+jets\xspace}

\newcommand{\ttW}        {\ensuremath{t \bar{t} + W}\xspace}
\newcommand{\ttZ}        {\ensuremath{t \bar{t} + Z}\xspace}
\providecommand{\ttbar}{blah}
\renewcommand{\ttbar}    {\ensuremath{t \bar{t}}\xspace}

\newcommand{\leptonic}   {{leptonic}}
\newcommand{\Leptonic}   {{Leptonic}}

\newcommand{\ptvec}      {{\ensuremath{\vec{p}_\mathrm{T}}}}

\newcommand{\akt}       {\ensuremath{{\textrm{anti-}}k_{t}}\xspace}

\newcommand{\neut}{\ensuremath{\tilde{\chi}_{1}^{0}}\xspace}
\newcommand{\neuttwo}{\ensuremath{\tilde{\chi}_{2}^{0}}\xspace}
\newcommand{\neutthree}{\ensuremath{\tilde{\chi}_{3}^{0}}\xspace}
\newcommand{\neuttwothree}{\ensuremath{\tilde{\chi}_{2,3}^{0}}\xspace}

\newcommand{\chargino}{\ensuremath{\tilde{\chi}_{1}^{\pm}}\xspace}

\newcommand{\rootgev}{\ensuremath{\,{\textrm{GeV}}^{1/2}}\xspace}
\newcommand{\metsig}{\ensuremath{\met/\sqrt{H_{\mathrm{T}}}}\xspace\xspace}

\newcommand{\Ht}{\ensuremath{H_{\textrm{T}}}\xspace}

\newcommand{\nfifty}{\ensuremath{{n_\mathrm{50}}}\xspace}
\newcommand{\neighty}{\ensuremath{{n_\mathrm{80}}}\xspace}
\newcommand{\nfiftyCR}{\ensuremath{{n_\mathrm{50}^\mathrm{CR}}}\xspace}
\newcommand{\neightyCR}{\ensuremath{{n_\mathrm{80}^\mathrm{CR}}}\xspace}
\newcommand{\nbjet}{\ensuremath{{n_{b\mathrm{-jet}}}}\xspace}
\newcommand{\multijet}{multijet}

\newcommand{\SR}[1]{\texttt{#1}\xspace}
\newcommand{\boldSR}[1]{\textbf{\SR{#1}}}

\providecommand{\ifb}{\ensuremath{fb^{-1}}}

\newcommand\Ptmiss{\ensuremath{\vec{E}_\mathrm{T}^{\mathrm{miss}}}\xspace}
\newcommand\etmissvec{\ensuremath{\vec{E}_\mathrm{T}^{\mathrm{miss}}}\xspace}

\newcommand{\bjet}{$b$-jet\xspace}
\newcommand{\bjets}{$b$-jets\xspace}

\newcommand{\btagged}{$b$-tagged\xspace}
\newcommand{\btagging}{$b$-tagging\xspace}

\newcommand{\madgraph}{\texttt{MadGraph}\xspace}
\newcommand{\powheg}{\texttt{Powheg}\xspace}
\newcommand{\powhegbox}{\texttt{Powheg-Box}\xspace}
\newcommand{\madspin}{\texttt{MadSpin}\xspace}

\newcommand{\herwigpp}{\texttt{Herwig++}\xspace}
\newcommand{\pythia}{\texttt{PYTHIA}\xspace}
\newcommand{\evtgen}{\texttt{EvtGen}\xspace}
\newcommand{\sherpa}{\texttt{SHERPA}\xspace}
\newcommand{\comix}{\texttt{Comix}\xspace}
\newcommand{\openloops}{\texttt{OpenLoops}\xspace}

\newcommand\paper{paper}
\newcommand\oursection[1]{\section{#1}}
\newcommand\oursubsection[1]{\subsection{#1}}

\AtlasTitle{Search for new phenomena in final states with large jet
multiplicities and missing transverse momentum with ATLAS using $\sqrt{s} =
13$~\TeV\ proton--proton collisions}

\author{The ATLAS Collaboration}

\AtlasAbstract{
Results are reported of a search for new phenomena, such as supersymmetric
particle production, that could be observed in high-energy proton--proton collisions.
Events with large numbers of jets,  together with missing transverse momentum from unobserved particles, are selected.
The data analysed were
recorded by the ATLAS experiment during 2015 using the
13\,\TeV~centre-of-mass proton--proton collisions at the Large Hadron Collider,
and correspond to an integrated luminosity of 3.2 fb$^{-1}$.
The search selected events with various jet multiplicities from $\ge7$ to $\ge 10$ jets,
and with various $b$-jet multiplicity requirements to enhance sensitivity.
No excess above Standard Model expectations is observed.
The results are interpreted within two supersymmetry models,
where gluino masses up to 1400\,\GeV~are excluded at 95\%\,confidence level,
significantly extending previous limits.
}

\AtlasJournal{Phys.\ Lett.\ B.}

\PreprintIdNumber{CERN-PH-EP-2016-026}

\AtlasRefCode{SUSY-2015-07}

\hypersetup{pdftitle={ATLAS draft},pdfauthor={The ATLAS Collaboration}}

\begin{document}

\oursection{Introduction}\label{sec:introduction}

New strongly interacting particles, if present at the \TeV~energy scale,
may be produced in high-energy proton--proton ($pp$) collisions and decay to final states with large jet multiplicities.
If their decay produces stable particles which only interact weakly,
it will also result in a momentum imbalance in the plane transverse to the beam (\etmissvec).

Such particles are present in supersymmetry (SUSY)~\cite{Golfand:1971iw,Volkov:1973ix,Wess:1974tw,Wess:1974jb,Ferrara:1974pu,Salam:1974ig},
a theoretically favoured extension of the Standard Model (SM)
that predicts partner fields for each of the SM particles.
These fields combine into physical superpartners of the SM particles. The
scalar partners of quarks and leptons are known as squarks ($\tilde{q}$) and
sleptons ($\tilde{\ell}$).
The fermionic partners of gauge and Higgs bosons are the gluinos ($\tilde{g}$),  the charginos ($\tilde{\chi}^{\pm}_{i}$, with $i$ = 1,2)
and the neutralinos ($\tilde{\chi}^{0}_{i}$ with $i$ = 1,2,3,4),
with $\tilde{\chi}^{\pm}_{i}$ and $\tilde{\chi}^{0}_{i}$ being the mass
eigenstates, ordered from the lightest to the heaviest, formed from the linear superpositions
of the SUSY partners of the Higgs and electroweak gauge bosons.

Under the
hypothesis of $R$-parity conservation~\cite{Farrar:1978xj}, SUSY partners are
produced in pairs and decay to the lightest supersymmetric particle (LSP),
which is stable and in a large variety of models is assumed to be the lightest
neutralino (\neut), which escapes detection. The undetected \neut would result in
missing transverse momentum, while the rest of the cascade can yield final
states with multiple jets and possibly leptons and/or photons.
The strongly interacting gluinos and squarks can have large production
cross-sections at
the Large Hadron Collider (LHC),
but no evidence of their existence has been observed to date.

This \paper{} presents the results of a search for new phenomena, such as supersymmetry,
in final states with large jet multiplicities (from $\ge$7 to $\ge$10 jets) in association with \met{}.
This signature is exhibited, for example, by squark
and gluino production followed by cascade decay chains,
and/or decays to heavy SM particles, such as top quarks or
$W$, $Z$ or Higgs bosons, each of which can produce multiple jets in their decays.
In contrast to many other searches for the production of strongly interacting SUSY particles,
the requirement made here of large jet multiplicity
means that the requirement on \met{} can be modest.

Previous searches~\cite{Aad:2011qa,Aad:2012hm,Aad:2013wta}
in similar final states have been performed by the ATLAS Collaboration
at the lower centre-of-mass energies of $\sqrt{s}=7\TeV$ and $8\TeV$,
with integrated luminosities up to $20.3\,$fb$^{-1}$.
The larger energy of the present dataset provides increased sensitivity,
particularly to particles with higher masses.
This paper closely follows the strategy of those previous studies.
In particular, data are collected using an online selection relying only on high jet multiplicity
and the signal regions (SR) are designed such that the
dominant \multijet{} background can be determined
from the data using regions of lower \met{} and/or lower jet multiplicity.

The data were collected by the ATLAS detector~\cite{Aad:2008zzm} in $pp$ collisions at the LHC
at a centre-of-mass energy of
$13\TeV$, from \nth{16}~August to \nth{3}~November 2015.
The detector covers the pseudorapidity\footnote{ATLAS uses a right-handed coordinate system with its origin at the nominal interaction point (IP)
in the centre of the detector and the $z$-axis along the beam pipe.  Cylindrical coordinates $(r,\phi)$ are
used in the transverse plane, $\phi$ being the azimuthal angle around the beam pipe.
The transverse momentum of a four-momentum is
$\ptvec = (p_{x}, p_{y})$,
its rapidity is $y=\frac{1}{2} \ln \frac{E+p_z}{E-p_z}$,
and the pseudorapidity is defined in terms of the polar angle~$\theta$ as $\eta=-\ln\tan(\theta/2)$.}
range of $|\eta| < 4.9$ and is hermetic in azimuth. It consists of an inner tracking detector surrounded by a
superconducting solenoid, electromagnetic and hadronic calorimeters, and an external muon spectrometer incorporating large superconducting toroidal magnets.
After applying beam-, data- and detector-quality criteria, the
integrated luminosity was
$3.2 \pm 0.2$\,fb$^{-1}$.
The uncertainty was derived using beam-separation scans,
following a methodology similar to that detailed in Ref.~\cite{Aad:2013ucp}.

\oursection{Physics object definition}\label{sec:objectselection}

Jets are reconstructed using the \akt{} clustering algorithm~\cite{Cacciari:2008gp,Cacciari:2005hq}
with jet radius parameter $R=0.4$ and
starting from clusters of calorimeter cells~\cite{Lampl:1099735}.
The effects of coincident $pp$ interactions (`pileup')
on jet energies are accounted for by an event-by-event $\pT$-density correction~\cite{ATLASpileup}.
The energy resolution of the jets is improved by using global sequential calibrations~\cite{ATLAS-CONF-2015-002,ATL-PHYS-PUB-2015-015}.
Events with jets originating from cosmic rays, beam background and detector noise are vetoed
using the `loose' requirements of Ref.~\cite{ATLAS-CONF-2015-029}.
Jets containing $b$-hadrons (\bjets)
are identified using an algorithm exploiting the
long lifetime, high decay multiplicity, hard fragmentation and large mass of $b$-hadrons~\cite{ATL-PHYS-PUB-2015-022}.
The \btagging{} algorithm tags \bjets with an efficiency of approximately 70\% in simulated \ttbar{} events,
and mis-tags $c$-jets, $\tau$-jets and light-quark or gluon jets with probabilities of approximately
10\%, 4\% and 0.2\% respectively~\cite{ATL-PHYS-PUB-2015-039}.

The primary vertex (PV) in each event is the vertex with the largest value of
$\Sigma p_{\mathrm{T}}^{2}$ for all tracks associated with it.
To reduce the effect of pileup,
a jet having $20\GeV<\pt<50\GeV$ and $|\eta|<2.4$ is disregarded when the
\pT-weighted sum of its associated tracks indicates that it originated from a pileup collision and not the PV,
based on a jet vertex tagger as described in Ref.~\cite{ATLASpileup}.

Electron candidates are identified according to the likelihood-based `loose'
criterion described in Ref. \cite{ATL-PHYS-PUB-2015-041},  formed from
e.g.~calorimeter shower shape and inner-detector track properties. Muon
candidates are identified according to the `medium' criterion described in
Ref.~\cite{ATL-PHYS-PUB-2015-037}, based on combined tracks from the
inner detector and muon spectrometer. These candidates (which may cause an
event to be rejected from the signal regions) are required to have
$\pT>10\GeV$, $|\eta|<2.47$ for $e$ and $|\eta|<2.5$ for $\mu$.

To avoid double-counting of reconstructed objects, electron candidates sharing
an inner-detector track with a muon candidate are removed. Next, jet candidates
separated from an electron candidate by \mbox{$\Delta R_y < 0.2$}
 are removed, where $\Delta
R_y = \sqrt{(\Delta y)^2 + (\Delta \phi)^2} $. Jet
candidates with fewer than three tracks and with $\Delta R_y <
0.4$ from a muon candidate are then removed. Following this, any lepton
candidate separated from a surviving jet candidate by $\Delta R_y < 0.4$ is removed.

The missing transverse momentum, \Ptmiss, is the negative two-vector
sum of the calibrated \ptvec{} of reconstructed jets with $\pt > 20\GeV$ and $|\eta| < 4.5$,
electrons, muons and photons~\cite{MET2015}. It includes an additional
contribution from inner-detector tracks, matched to the PV,
that are not associated with these reconstructed objects.
Photons are not considered beyond their contribution to the \Ptmiss unless they are reconstructed as jets.
To reduce the effect of pileup, jets do not contribute to the \Ptmiss{} calculation
when they are disregarded based on the jet vertex tagger as described
above.
Additionally, when a jet having $50\GeV< \pT < 70\GeV$, $|\eta|<2.0$ and
azimuth relative to the missing momentum $\Delta\phi(\vec{p}_{\mathrm T},\,\Ptmiss) > 2.2$
meets the same vertex-tagging criterion, the event is discarded. Events
in which the jet closest in $\phi$ to the \Ptmiss is found in or near an inactive region
in the hadronic calorimeter barrel (i.e. $-0.1 < \eta < 1.0$, $0.8 <
\phi < 1.1$) are also discarded, in order to reduce the impact of this source of \Ptmiss mismeasurement.
These data-quality requirements reduce the expected acceptance of typical SUSY models by approximately 5\%.

When defining leptons for control regions (Section~\ref{sec:systematics}),
the candidates defined above are required to be isolated, to have a longitudinal impact parameter $z_0$ (with respect to the PV) satisfying
$|z_{0}\sin\theta| < 0.5$\,mm, and to have the significance of their transverse impact parameter $|d_{0}/\sigma(d_{0})|$ (with respect to the measured beam position)
be less than five for electrons and less than three for muons.
Additionally, electrons must satisfy the `tight' criterion of Ref. \cite{ATL-PHYS-PUB-2015-041}.

\begin{table}[b]
\centering
\renewcommand\arraystretch{1.35}
\addtolength{\tabcolsep}{-1pt}
\subfigure[Signal regions using \nfifty{}.]{
\begin{small}
\hspace*{-2mm}\begin{tabular}{|l||c|c|c||c|c|c||c|c|c|}
\hline
               &\boldSR{8j50} & \boldSR{8j50-1b} & \boldSR{8j50-2b}
               & \boldSR{9j50} & \boldSR{9j50-1b} & \boldSR{9j50-2b}
               & \boldSR{10j50} & \boldSR{10j50-1b} & \boldSR{10j50-2b} \\
\hline
\nfifty  & \multicolumn{3}{c||}{$\geq 8$}   & \multicolumn{3}{c||}{$\geq 9$} & \multicolumn{3}{c|}{$\geq 10$} \\
\hline
\nbjet   & ---  & $\geq 1$ & $\geq 2$       &  ---  & $\geq 1$ & $\geq 2$    & ---  & $\geq 1$ & $\geq 2$ \\
\hline
\metsig  & \multicolumn{9}{c|}{$> 4\rootgev$}\\
\hline
\end{tabular}\end{small}
\
}
\addtolength{\tabcolsep}{+1pt}
\subfigure[Signal regions using \neighty{}.]{
\begin{small}
\begin{tabular}{|l||c|c|c||c|c|c|}
\hline
               & \boldSR{7j80} & \boldSR{7j80-1b} & \boldSR{7j80-2b}
               & \boldSR{8j80} & \boldSR{8j80-1b} & \boldSR{8j80-2b}\\
\hline
\neighty  & \multicolumn{3}{c||}{$\geq 7$}   & \multicolumn{3}{c|}{$\geq 8$} \\
\hline
\nbjet   & ---  & $\geq 1$ & $\geq 2$       &  ---  & $\geq 1$ & $\geq 2$    \\
\hline
\metsig  & \multicolumn{6}{c|}{$> 4\rootgev$}\\
\hline
\end{tabular}\end{small}
\
}
\caption{\label{tab:SRdefs}%
Definition of the signal regions.
The selection variables are described in
Sections~\ref{sec:objectselection} and~\ref{sec:eventselection}.
A long dash `---' indicates that no requirement is made.
Events with leptons are vetoed.
}
\end{table}

\oursection{Event selection}\label{sec:eventselection}

The signal regions are defined using two jet multiplicity counts:
either \nfifty{}, the number of jets having $\pt>50\GeV$ and $|\eta|<2.0$,
or \neighty{}, the number of such jets which additionally satisfy the higher
requirement $\pt>80\GeV$.
The online selection (trigger) for \nfifty{}-based regions requires events to have at least six jets each
with $\pt > 45\GeV$ and $|\eta| < 2.4$, while that for \neighty{}-based regions requires at least five jets each with $\pt > 70$\,\GeV.
The trigger efficiency is greater than 99.5\% for events
satisfying the signal selection described below.
Jets with a looser definition -- those having $\pt>40\GeV$ and  $|\eta|<2.8$ --
are used to construct the scalar sum $\Ht = \sum \pt^{\mathrm{jet}}$,
while those having $\pt>40\GeV$ and $|\eta|<2.5$ are candidates for \btagging{},
contributing to the number \nbjet{} of $b$-tagged jets.

The signal selection requires large jet multiplicity,
which depends on the signal region (SR), as shown in Table~\ref{tab:SRdefs}.
Fifteen different SRs are defined, providing wide-ranging sensitivity to models
with different final states and mass spectra.
There are three different triplets of regions defined in terms of
the jet multiplicity \nfifty{}
and two different triplets of regions defined in terms of \neighty{}.
Within each triplet, different requirements are made on \nbjet{},
from no requirement to the requirement of at least two \bjets.
In all cases the final selection is on the ratio of \met{} to $\sqrt{\HT}$,
with the choice of a threshold at 4\rootgev being a good balance between background rejection and signal efficiency
while maintaining the effectiveness of the background estimation.
Events containing electron or muon candidates
with $\pt>10\GeV$ are vetoed to reduce background from SM processes.

The SRs have events in common, for example all events in \SR{9j50-1b}
 also appear in \SR{9j50}, which does not require the \bjet,
and in \SR{8j50} and \SR{8j50-1b}, which have a looser requirement on \nfifty{}.
Events may also appear in both the \nfifty{} and the \neighty{}
categories.

\oursection{Background and simulation}\label{sec:bg}

Standard Model processes contribute to the event counts in
the SRs. The dominant background contributions are \multijet{}
production, including those from purely strong interaction processes
and fully hadronic decays of \ttbar{}; partially leptonic
decays of \ttbar; and leptonically decaying $W$ or $Z$ bosons produced
in association with jets.
Top-quark, $W$- and $Z$-boson decays that are not fully hadronic
are collectively referred to as `leptonic' backgrounds. They can
contribute to the signal regions when no $e$ or $\mu$ leptons are produced, for
example $Z\to\nu\nu$ or hadronic $W \to \tau\nu$ decays, or
when they are produced but are out of acceptance, lie within jets, or are not reconstructed.

The most significant \leptonic{} backgrounds
are \ttbar and $W$ boson production in association with jets.
The contribution of these two backgrounds to the signal regions
is determined from a combined fit
as described later in Section~\ref{sec:fit}.
The yields for the other, generally subdominant, \leptonic{} backgrounds are taken
from the simulations as described below.

Monte Carlo simulations are used in the determination of the \leptonic{} backgrounds
and to assess sensitivity to specific SUSY signal models.
All simulated events are overlaid with multiple $pp$ collisions
simulated with the soft QCD processes of \pythia 8.186 \cite{pythia8.1} using the A2 set of parameters (tune)~\cite{ATLAS:2012uec}
and the MSTW2008LO parton distribution functions (PDF) \cite{Martin:2009iq}.
The simulations are weighted such that the pileup conditions match those of the data.
The response of the detector to particles is modelled with an ATLAS detector simulation \cite{:2010wqa}
based fully on \textsc{Geant4} \cite{Agostinelli:2002hh}, or using fast
simulation based on a parameterisation of the performance
of the ATLAS electromagnetic and hadronic calorimeters \cite{atlfast} and on \textsc{Geant4} elsewhere.
\Leptonic{} background samples use full simulation, while signal samples (described below) use the fast simulation option.
Corrections are applied to the simulated samples to account for differences between data and simulation for the lepton identification
and reconstruction efficiencies, and for the efficiency and misidentification rate of the \btagging algorithm.

\oursubsection{Leptonic background simulation}\label{sec:bg:leptonic}

For the generation of \ttbar and single top quarks in the $Wt$ and $s$-channels \cite{ATLAS:simul:top}
\powhegbox~v2~\cite{powheg-box} is used with the CT10 PDF sets \cite{CT10pdf} in the matrix element calculations.
Electroweak $t$-channel single-top-quark events are generated using \powhegbox~v1.
This generator uses the four-flavour scheme for the next-to-leading order (NLO) matrix element calculations
together with the fixed four-flavour PDF set CT10f4 \cite{CT10pdf}.
For this process, the top quarks are decayed using \madspin~\cite{madspin} preserving all spin correlations,
while for all processes  the parton shower, fragmentation,
and the underlying event are simulated using \pythia~v6.428~\cite{pythia6} with the CTEQ6L1 PDF sets \cite{Pumplin:2002vw}
and the corresponding Perugia 2012 tune (P2012) \cite{perugia}.
The top quark mass is set to $172.5\GeV$.
The \evtgen~v1.2.0 program~\cite{evtgen} models the bottom and charm hadron decays, as it does for all non-\sherpa-simulated processes mentioned below.
The \ttbar{} simulation is normalised to the cross-section calculated to
next-to-next-to-leading order
(NNLO) in perturbative QCD,
including soft-gluon resummation to next-to-next-to-leading-log (NNLL) accuracy
\cite{top++}.

Events containing \ttbar and additional heavy particles -- comprising three-top,
four-top, \ttW, \ttZ and $\ttbar + WW$ production \cite{ATLAS:simul:ttV} --
are simulated at leading order in the strong coupling constant
$\alpha_\mathrm{s}$,
using \madgraph~v2.2.2 \cite{Alwall:2014hca} with up to two additional partons in the matrix element,
interfaced to the \pythia~8.186 \cite{pythia6,pythia8.1}  parton shower model.
The A14 tune of the \pythia parameters is used~\cite{a14}, together with the NNPDF2.3LO PDF set~\cite{NNPDF}.
The predicted production cross-sections are calculated to NLO as described in Ref.~\cite{Alwall:2014hca}
for all processes other than three-top, for which it is calculated to LO.

Events containing $W$ bosons or $Z$ bosons with associated jets
\cite{ATLAS:simul:V+jets} are likewise simulated using \madgraph, but with up
to four additional final-state partons in the matrix element, and interfaced to
\pythia, using the same tunes and particle decay programs. The $W$ + jets and $Z$ +
jets events are normalised to NNLO cross-sections \cite{Gavin:2010az}.
Diboson processes with at least one boson decaying leptonically
\cite{ATLAS:simul:VV} are simulated using the \sherpa~v2.1.1
generator~\cite{sherpa2}. The matrix element calculations contain all diagrams
with four electroweak vertices. They are calculated for up to one (for 4$\ell$,
2$\ell$+2$\nu$, semileptonic $ZZ$) or no additional partons (for
3$\ell$+1$\nu$, other semileptonic processes) at NLO and up to three additional partons
at LO using the \comix \cite{comix} and \openloops \cite{openloops} matrix
element generators and interfaced with the \sherpa parton shower
\cite{sherpashower} using the ME+PS@NLO prescription \cite{mepsnlo}. The CT10
PDF set is used in conjunction with dedicated parton shower tuning developed by
the \sherpa authors.

Theoretical uncertainties are considered for all these simulated samples.
Production of \ttbar is by far the most important process simulated in this
analysis and to evaluate the uncertainty on this background several samples are
compared. Samples are produced with the factorisation and renormalisation
scales varied coherently, along with variations of the resummation damping
parameter and with more/less radiation tunes of the parton shower
\cite{ATL-PHYS-PUB-2015-011}. Additionally the nominal sample is compared to
one with \powheg interfaced with \herwigpp
 \cite{Bahr:2008pv} and \sherpa~v2.1.1
samples with up to one additional jet at next-to-leading order using \openloops
and up to four additional jets at leading order, to account for uncertainties
in the parton shower and the generator respectively. The comparison with the
\sherpa sample dominates the uncertainty in the signal region prediction.

\oursubsection{SUSY signal models}\label{sec:signals}
Two classes of SUSY signal models are used when interpreting the results.
The first is a simplified model, in which gluinos are pair-produced and then decay
via the cascade
\begin{align*}
\gluino  & \to q + \bar{q}' + \chargino \hspace*{0.3cm} (q = u,d,s,c) \\
\chargino    & \to W^{\pm} + \neuttwo \\
\neuttwo & \to Z + \neut.
\end{align*}
The parameters of the model are the masses of the gluino, $m_{\gluino}$, and the lightest neutralino, $m_{\neut}$.
The mass of the $\chargino$ is constrained to be $\frac{1}{2} (m_{\gluino} + m_{\neut})$
and the mass of the $\neuttwo$ to be  $\frac{1}{2} (m_{\chargino} + m_{\neut})$.
All other sparticles are kinematically inaccessible.
This model is labelled in the following figures as `2-step'.

A second set of SUSY models is drawn from a two-dimensional subspace (a `slice') of the
19-parameter phenomenological Minimal Supersymmetric Standard Model (pMSSM)~\cite{Djouadi:1998di,Berger:2008cq}. The selection is motivated in part by
models not previously excluded in the analysis presented in Ref.~\cite{Aad:2015baa}.
The models are selected to have a bino-dominated \neut{}, kinematically accessible gluinos,
and a Higgsino-dominated multiplet at intermediate mass.
The Higgsino multiplet contains two neutralinos (the \neuttwo{} and \neutthree{}) and a chargino (the \chargino).
The mass of these particles is varied by changing the soft SUSY-breaking parameters
$M_3$ (for the gluino), $M_1$ (for the \neut, set to $60\GeV$), and $\mu$ (for the Higgsinos).
In order that other SUSY particles remain kinematically inaccessible, the other parameters, defined in Ref.~\cite{Aad:2015baa}, are set to
$M_A=M_2=3\TeV,$ $A_\tau=0$, $\tan\beta=10$,
$A_t=A_b=m_{\tilde{L}_{(1,2,3)}}=m_{(\tilde{e},\tilde{\mu},\tilde{\tau})}=m_{\tilde{Q}_{(1,2,3)}}=m_{(\tilde{u},\tilde{c},\tilde{t})}=m_{(\tilde{d},\tilde{s},\tilde{b})}=5\TeV$.
Mass spectra with consistent electroweak symmetry breaking
are generated using \texttt{softsusy}~3.4.0~\cite{Allanach:2001kg}.
The decay branching ratios are calculated with \texttt{SDECAY/HDECAY}~1.3b/3.4~\cite{Djouadi:2006bz},
and when $m_{\chargino} \lsim 500\GeV$ and $m_{\gluino} \gsim 1200\GeV$ the predominant decays are
$\gluino  \to t + \bar{t} + \neuttwothree$ and $\gluino  \to t + \bar{b} + \chargino$, with \neuttwothree
decaying to $Z / h + \neut$ and \chargino to $W^{\pm} + \neut$ (numerical values are provided in Ref. \cite{hepdata}).
When these decays dominate they lead to final states with many jets, several of which are \bjets, but relatively little \met.
This renders this search particularly sensitive compared to most other SUSY searches, which tend to require high \met.
At higher $m_{\chargino}$ and lower $m_{\gluino}$, the decay $\gluino \to q + \bar{q} + \neut$ becomes dominant and this search starts to lose sensitivity.
This model is labelled in the following figures as `pMSSM'.

The signal events are simulated
using \madgraph~v2.2.2 at LO interfaced to \pythia~8.186,
as for those of \wjets and \zjets.
The signal cross-sections are calculated at NLO in the
strong coupling constant, adding the resummation of soft gluon emission at
next-to-leading-logarithmic (NLL) accuracy \cite{Beenakker:1996ch,Kulesza:2008jb,Kulesza:2009kq,Beenakker:2009ha,Beenakker:2011fu}.
The nominal cross-section is taken from an envelope of
cross-section predictions using different PDF sets and factorisation and
renormalisation scales, as described in Ref.~\cite{Kramer:2012bx}.

\begin{figure}\begin{center}
\hfill
\subfigure[$\nfifty=7$, using a template with $\nfifty=6$.]{
\includegraphics[width=0.49\textwidth]{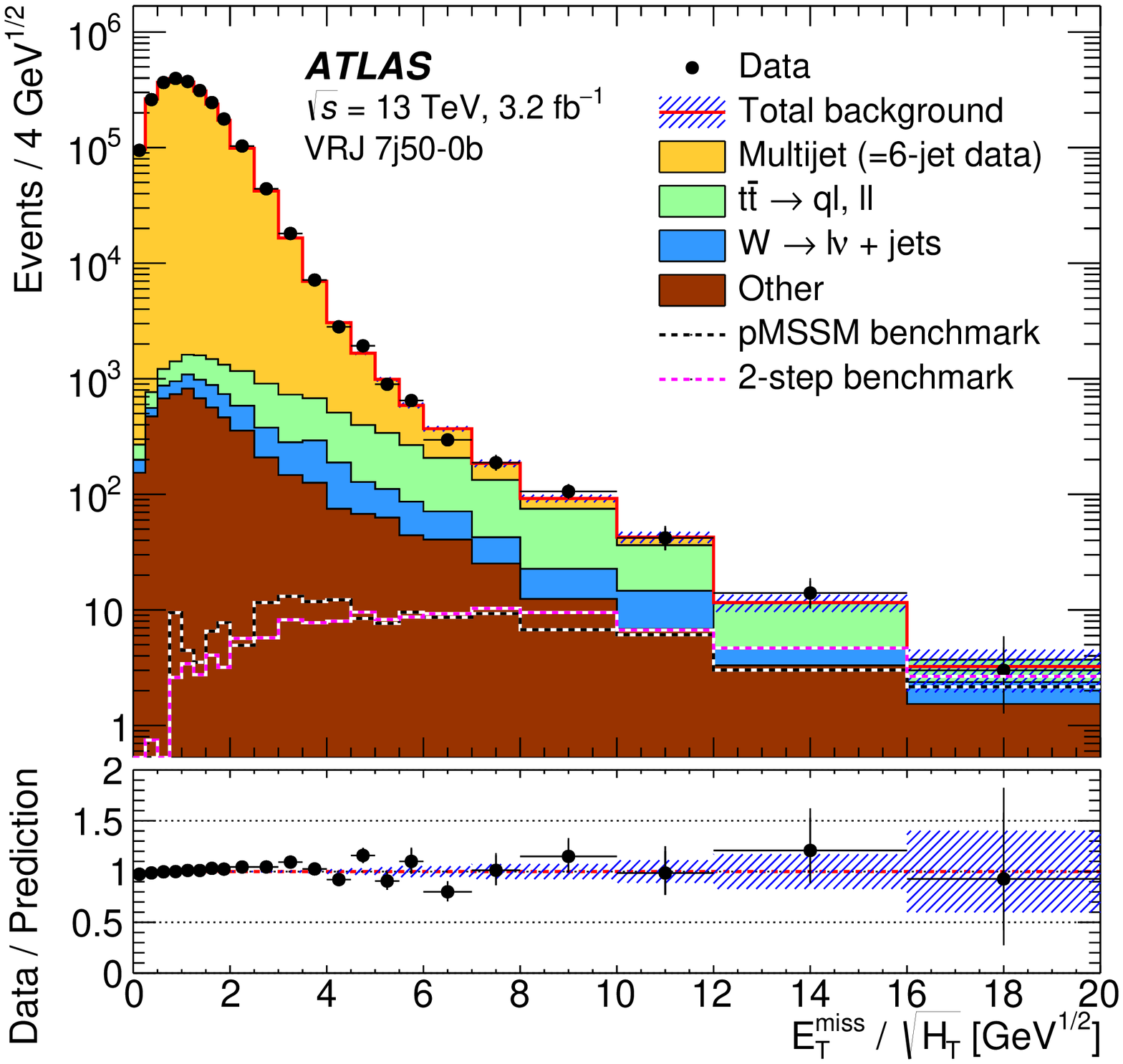}}
\hfill
\subfigure[$\neighty=6$, using a template with $\neighty=5$.]{
\includegraphics[width=0.49\textwidth]{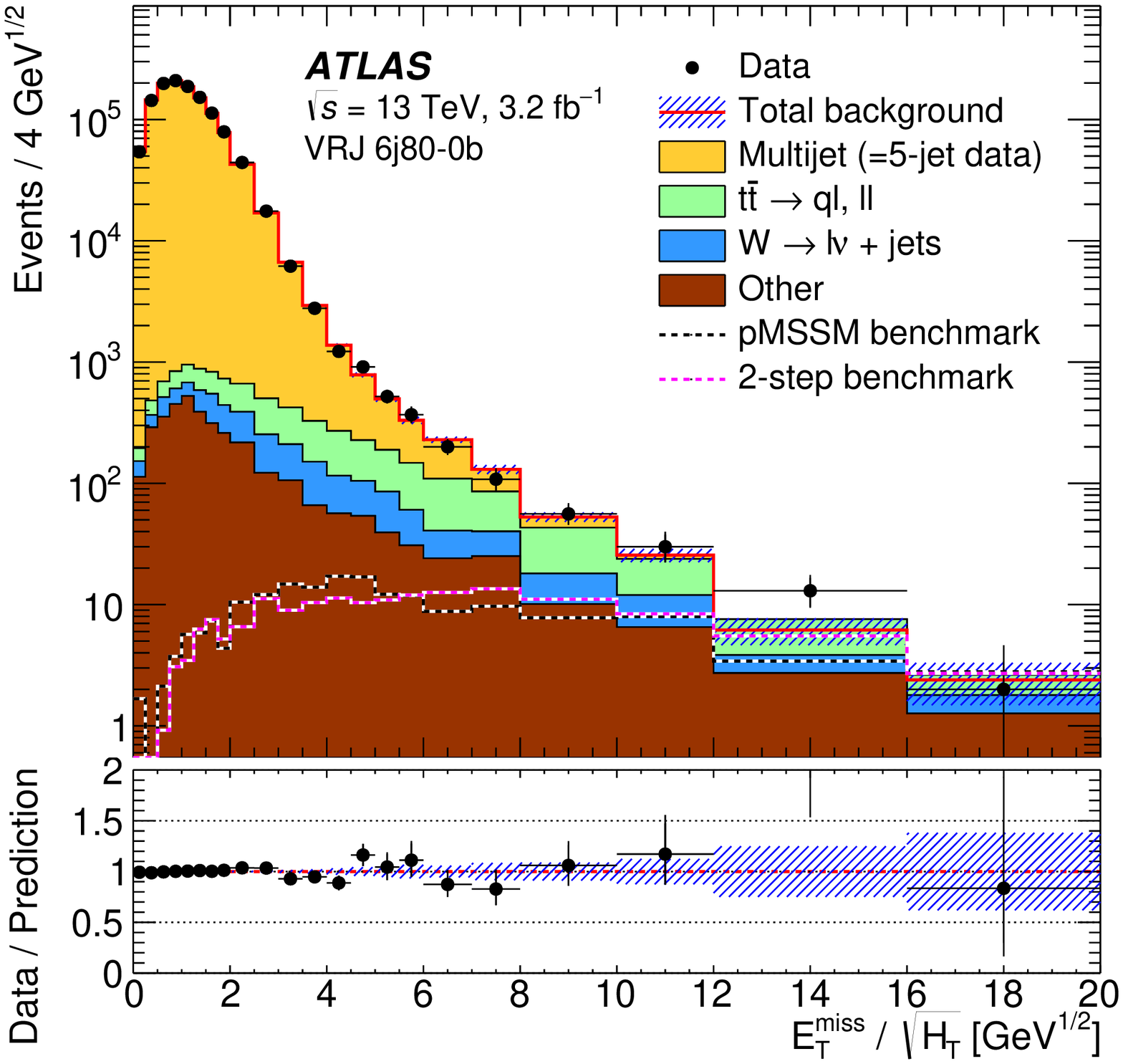}}
\hfill\\
\caption{\label{fig:multijet_templates}
Example distributions of the ratio \metsig{}
in the validation region
with jet multiplicities (a) $\nfifty=7$, and (b) $\neighty=6$.
The templates, which are in each case built from a jet multiplicity one
smaller than that of the data,
are normalised to the data in the region with $\metsig<1.5\rootgev$.
The templates are weighted by \HT{} as described in the text.
No requirement is made on $\nbjet$.
Variable bin sizes are used with bin widths (in units of \rootgev) of 0.25 (up to $\metsig = 2$), 0.5 (from 2 to 6), 1 (from 6 to 8), 2 (from 8 to 12) and 4 thereafter, with the last bin additionally containing all events with $\metsig > 20\rootgev$.{}
The total background can lie below the \leptonic{} background contribution in individual bins,
since the template can give a negative contribution.
The dashed lines labelled `pMSSM' and `2-step'
show the (small) signal contamination from example SUSY models
described in Section~\ref{sec:signals} -- a pMSSM model point with
($m_{\gluino}$,~$m_{\chargino}$) = (1300,~200)\,\GeV~and
a cascade decay model with ($m_{\gluino}$,~$m_{\neut}$) = (1300,~200)\,\GeV.
The sub-plots show the ratio of the data to the SM prediction, with the blue hatched
band showing the statistical uncertainty arising from a finite number of MC events and
limited data in the templates and $\metsig<1.5$ normalisation regions.
}
\end{center}\end{figure}

\begin{figure}
\centering
\subfigure[$\nbjet=0$]     {\includegraphics[width=0.49\textwidth]{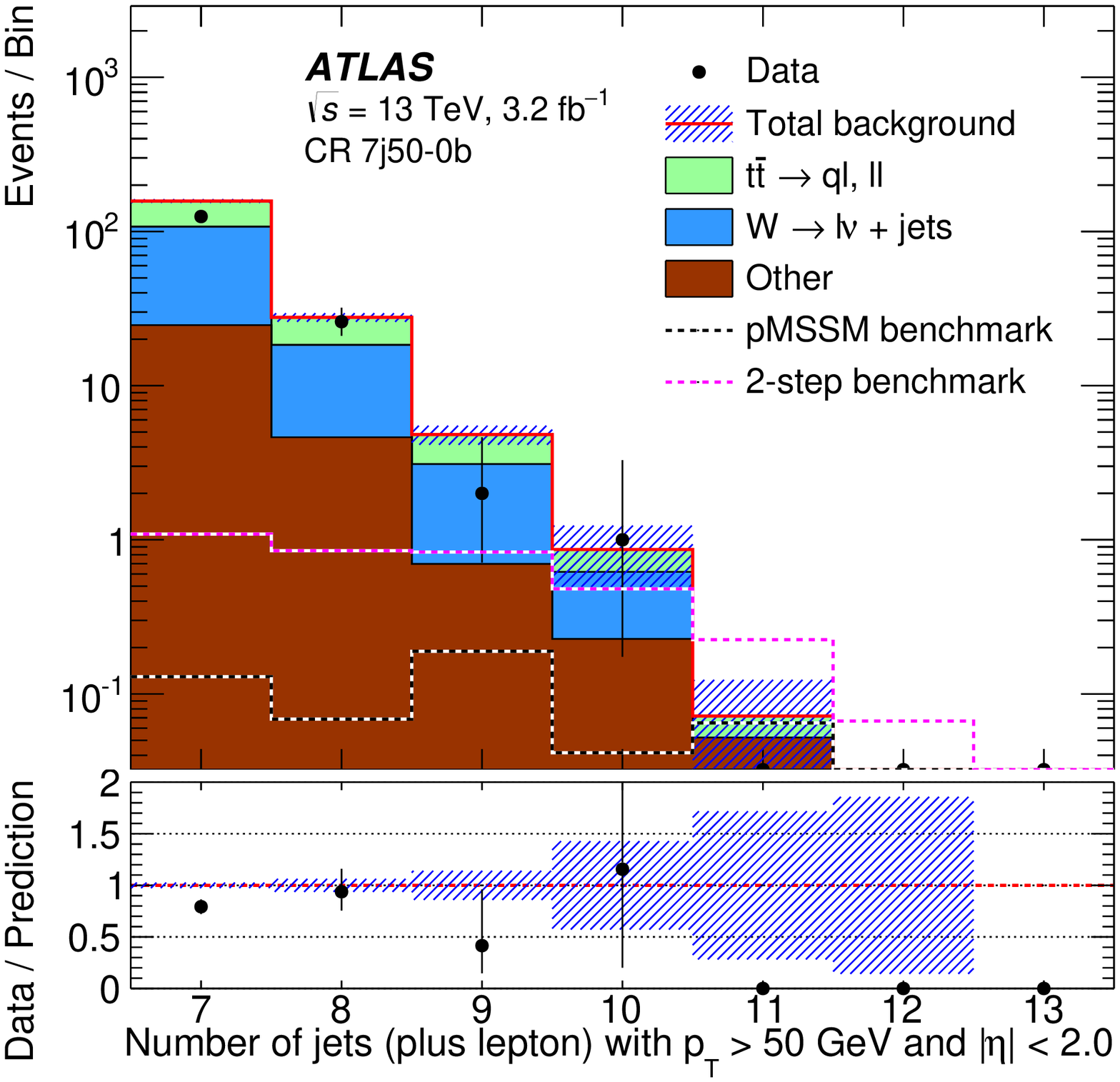}}
\subfigure[$\nbjet \geq 1$]{\includegraphics[width=0.49\textwidth]{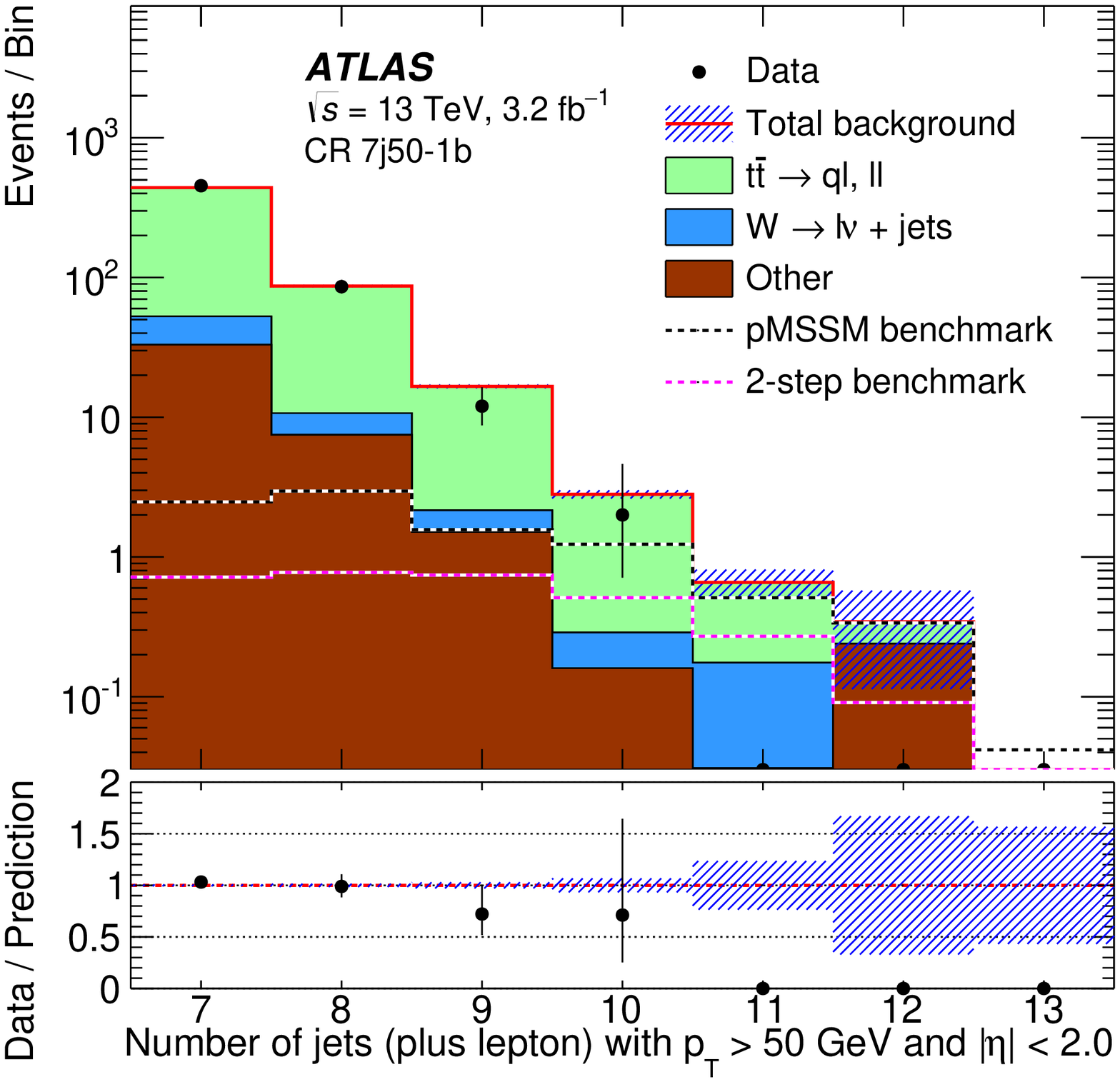}}
\caption{
Control regions -- requiring one lepton -- showing the \nfifty{} jet multiplicity distributions
after all selections aside from \nfifty{}.
That lepton is permitted to contribute to the jet multiplicity count and to \HT{}.
The sub-plots show the ratio of the data to the Standard Model prediction.
The blue hatched bands on those sub-plots show MC statistical uncertainties.
The dashed lines labelled `pMSSM' and `2-step' refer to
benchmark signal points -- a pMSSM slice model with ($m_{\gluino}$,
$m_{\chargino}$) = (1300, 200)\,\GeV~and
a cascade decay model with ($m_{\gluino}$, $m_{\neut}$) = (1300, 200)\,\GeV.
All backgrounds are normalised according to their theoretical (pre-fit) cross-sections.
}
\label{fig:CR:njet}
\end{figure}

\begin{figure}\begin{center}
\subfigure[$\nfifty\ge10$.]{
\includegraphics[width=0.49\textwidth]{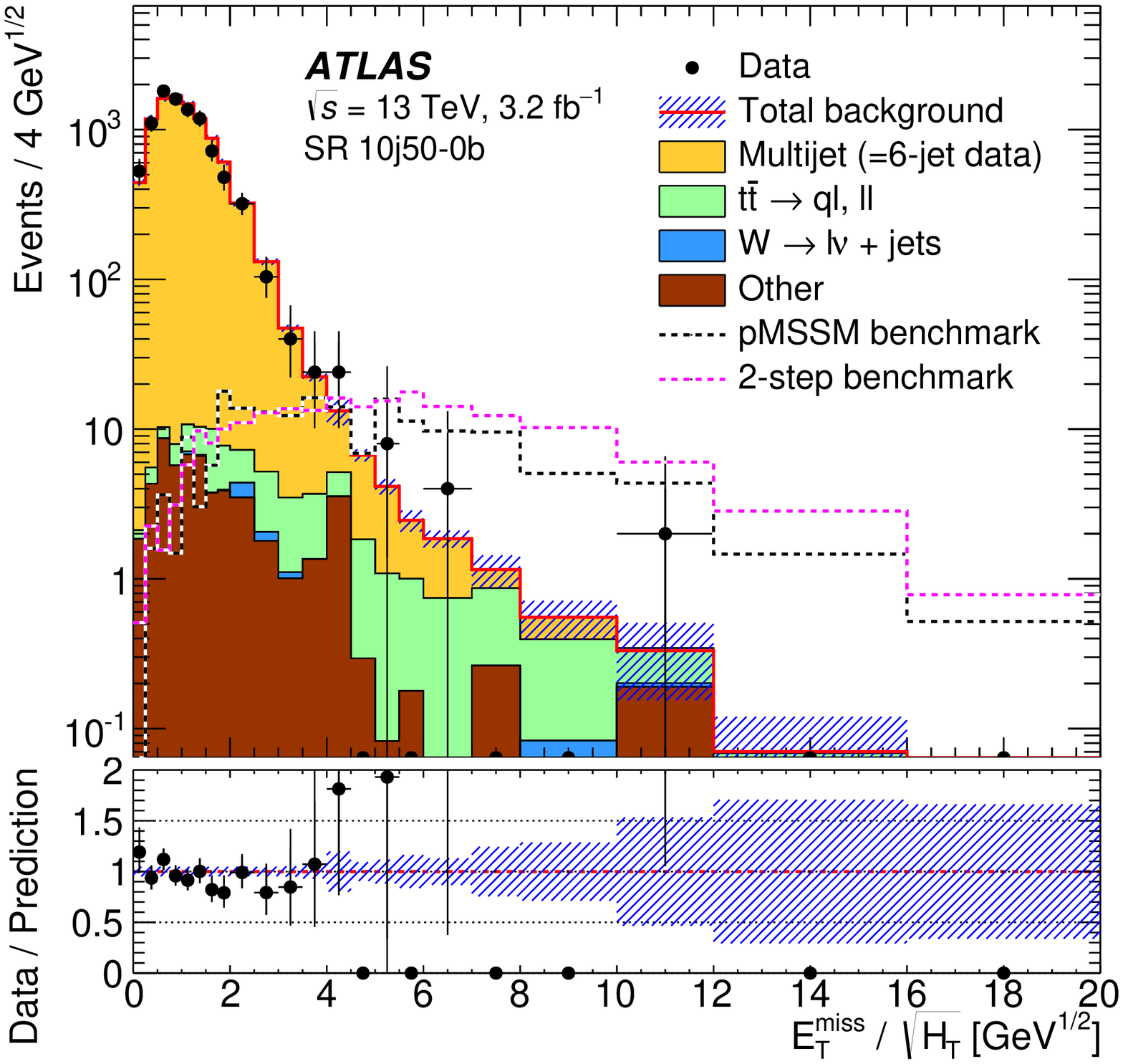}}
\subfigure[$\nfifty\ge10$ and $\nbjet\geq2$.]{
\includegraphics[width=0.49\textwidth]{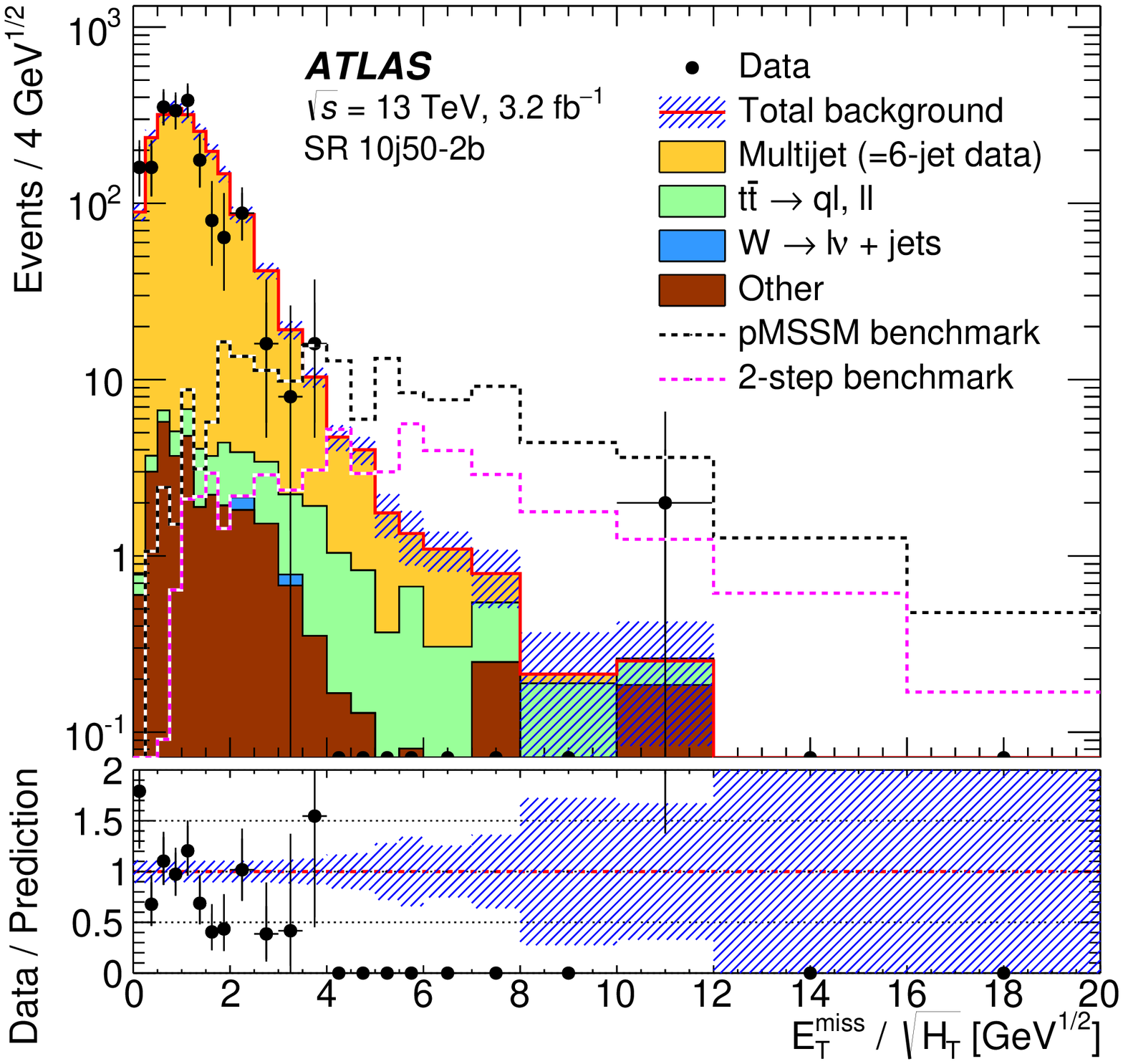}}
\subfigure[$\neighty\ge8$.]{
\includegraphics[width=0.49\textwidth]{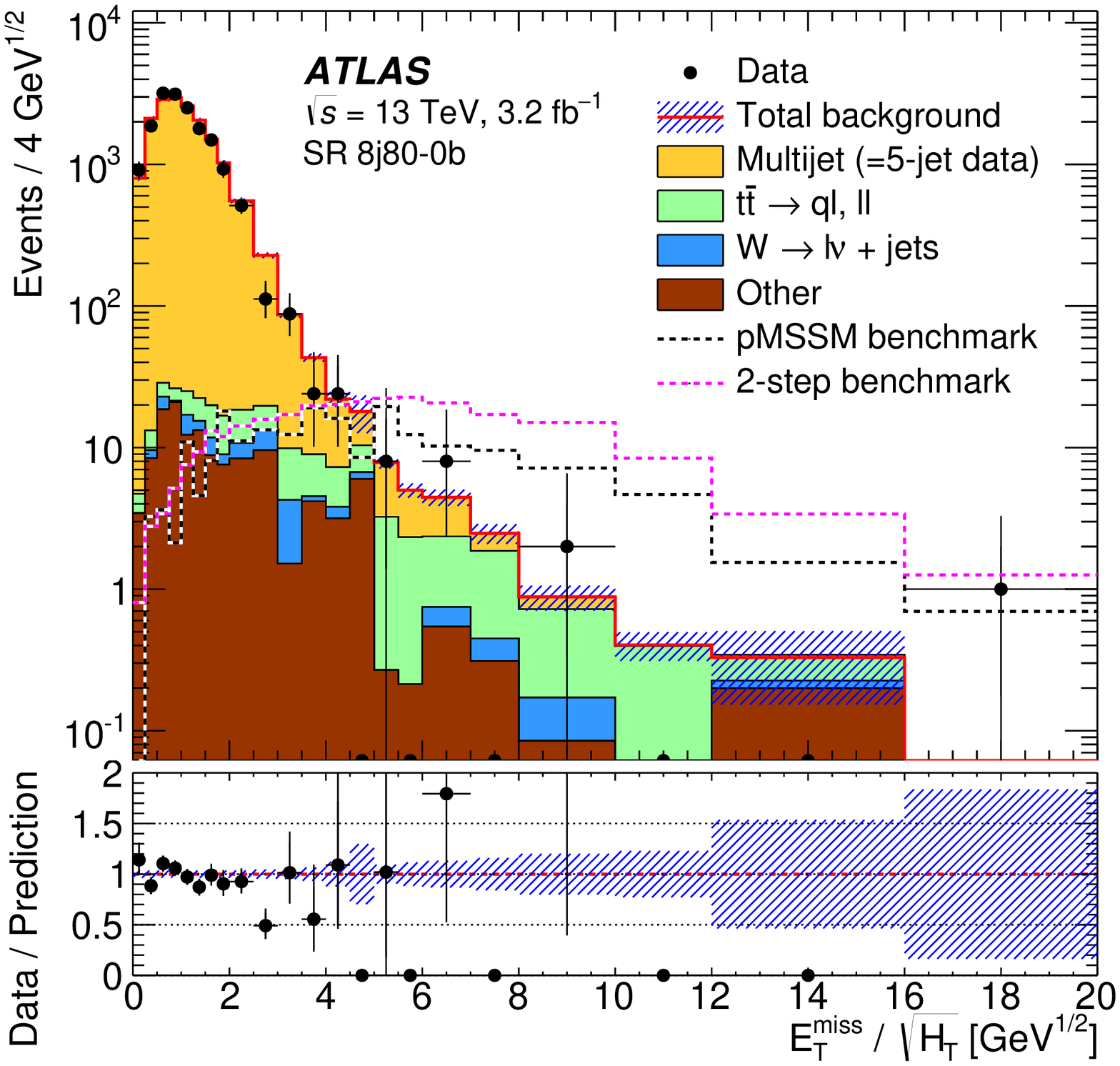}}
\subfigure[$\neighty\ge8$ and $\nbjet\geq2$.]{
\includegraphics[width=0.49\textwidth]{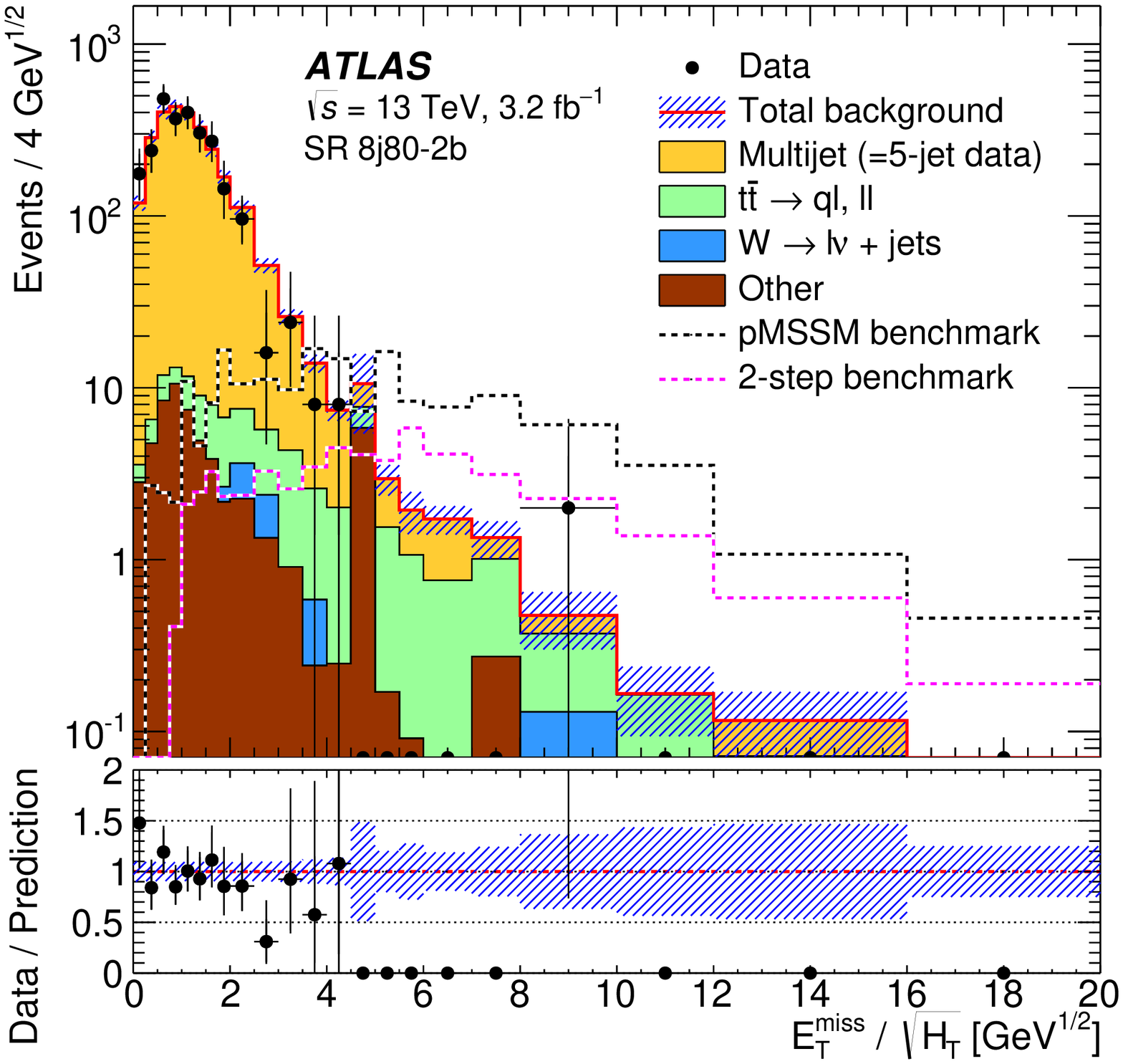}}
\caption{\label{fig:metsig_final}
Example distributions of the selection variable \metsig{}, for the largest multiplicities
required of the number of jets with \pt{} larger than 50\,\GeV{} (top) or 80\,\GeV{} (bottom).
The plots on the left have no selection on the number of $b$-tagged jets, while those on the right are for events with $\nbjet\geq2$.
\wjets and \ttbar are normalised to their post-fit values, while the other \leptonic{} backgrounds are normalised to their theoretical cross-sections.
The \multijet{} templates are normalised to data at lower jet multiplicities
in the region $\metsig{}<1.5\rootgev$, in the manner described in Section~\ref{sec:bg:multijet}.
The SRs lie where $\metsig{}>4\rootgev$.
The dashed lines labelled `pMSSM' and `2-step' refer to
benchmark signal points -- a pMSSM slice model with ($m_{\gluino}$,
$m_{\chargino}$) = (1300, 200)\,\GeV~and
a cascade decay model with ($m_{\gluino}$, $m_{\neut}$) = (1300, 200)\,\GeV.
Other details are as described in Figure~\ref{fig:multijet_templates}.
}
\end{center}\end{figure}

For the model points shown later in Figures \ref{fig:multijet_templates}--\ref{fig:metsig_final},
with $m_{\gluino}=1300\GeV$ slightly beyond the Run-1 exclusion limits,
the SR selection efficiencies are around 8\% in the SRs most sensitive to those models.

\oursubsection{Multijet background}\label{sec:bg:multijet}

The signal regions were chosen such that the background
from the \multijet{} process can be determined from the data.
The method relies on the observation~\cite{Aad:2011qa} that where \met{}
originates predominantly from calorimeter energy mismeasurement,
as is the case for the \multijet{} contributions,
the distribution of the ratio \metsig{} is almost invariant under changes in jet multiplicity.
This invariance, which is illustrated in Figure~\ref{fig:multijet_templates}, occurs because the calorimeter resolution that produces the momentum imbalance
in these events is dominated by stochastic processes which have variance proportional to \HT{},
and is largely independent of the jet multiplicity.

\newcommand{\plotdir}{WillKplotdump/SUSY4_12Jan_speedy}
\newcommand{\bonusprelim}{_Internal}
\renewcommand{\bonusprelim}{}
\newcommand{\HTbinning}{_HT_0-600-800-1000-inf\bonusprelim_mod}
\newcommand{\binwidths}{Variable bin sizes are used with bin widths (in units of \rootgev)
of 0.25 (up to $\metsig = 2$), 0.5 (from 2 to 6), 1 (from 6 to 8), 2 (from 8 to 12) and 4 thereafter,
with the last bin additionally containing all events with $\metsig > 20\rootgev$.}

The shape of the \metsig{} distribution is measured
in control regions (CR) with lower jet multiplicities than the signal regions,
and correspondingly much higher multijet contributions.
For the \nfifty{} signal regions,
the CR contains events with exactly six jets having $\pt>50\GeV$.
For the \neighty{} signal regions, the CR requires exactly five jets
with $\pt>80\GeV$.
For each SR jet selection, an appropriate \metsig{} distribution template is normalised to the data
in a further CR having the same jet multiplicity as the SR but with $\metsig<1.5\rootgev$.
That normalised template then provides the background prediction for
the SR multiplicity in the region with $\metsig>4\rootgev$.

Since semileptonic $b$-hadron decays can contribute to \met{},
these \metsig{} template distributions are built separately for each $\nbjet$ requirement.
For example, the \multijet{} contribution to the \SR{9j50-1b}
signal region is determined using a template built from
events with exactly six jets with $\pt>50\GeV$, and $\nbjet \ge 1$.
That template is normalised to \SR{9j50-1b} in the region with
$\metsig<1.5\rootgev$.

When constructing and normalising the \metsig{} templates,  the same lepton
veto is used as for the signal regions.
However, some \leptonic{} background contributions persist,
and so the expected \leptonic{} backgrounds to those templates
(normalised according to their theoretical cross-sections, as described in Section~\ref{sec:bg:leptonic})
are subtracted from the data distributions. The
uncertainties associated with the \leptonic{} backgrounds are included in the
systematic uncertainty in the prediction. Non-stochastic contributions to
calorimeter resolution, which lead to a  residual dependence of the \metsig{}
distribution on \HT{}  (at the $\mathcal{O}(10\%)$ level), are reduced by
constructing the templates in four bins of  \HT{} in the kinematic region of
interest.  Those proto-templates are combined with weights which reflect the
\HT{} distribution of the CR with the same jet multiplicity as the target SR
but with $\metsig<1.5\rootgev$. The effect of changing the \HT binning is
included in the systematic uncertainty.

The validity of assuming \metsig{} invariance is tested with data,
using a series of validation regions (VR) with smaller jet multiplicities
or smaller \metsig{} (between $1.5\rootgev$ and $3.5 \rootgev$) than the SRs, or both.
These VRs are found to be described by the templates,
constructed as described above, mostly to within 10\%--20\%.
However, for the tightest regions (with very few events) the discrepancy reaches 60\%.
The tests are performed separately for each of the three
$b$-jet requirements, and the largest difference for each set,
including VRs with jet multiplicity up to and including that of the SR in question,
is included as an overall `closure' systematic uncertainty associated with the method.

\oursection{Statistical treatment and systematic uncertainties}\label{sec:systematics}

Systematic uncertainties specific to the \multijet{} and \leptonic{}
background contributions are described in Sections~\ref{sec:bg:multijet} and \ref{sec:bg:leptonic}
respectively.
Further uncertainties that apply to signal processes and all simulated backgrounds include
those on the jet energy scale, jet resolution,
integrated luminosity, the \btagging efficiency (for correct and incorrect
identifications of both the $b$- and non-$b$-jets),
and the lepton identification efficiency and energy scale.
They are in general small compared to the aforementioned ones, being at most one third the size of the largest of those.

The effect of the systematic uncertainties on the SM background calculations
is reduced by constraining the normalisations of the \ttbar{} and $W$+jets backgrounds
using dedicated control regions kinematically close to, but distinct from, the signal regions, as shown in Table~\ref{tab:CRdefs}.
Each \leptonic{} control region contains events with one electron or muon that meets the stricter requirements described in Section~\ref{sec:objectselection}
and has transverse momentum $\pT^\ell>20\GeV$.
There must be no additional lepton candidates with $\pT^\ell>10\GeV$.
Each such region uses the same multijet trigger as its corresponding SR.

To reduce generic background from new particles which may decay to a final state with leptons and \met,
a modest upper bound of 120\,\GeV\ is placed on the transverse mass
$\mt = \left(2\, \met\pt^\ell - 2\,\Ptmiss\cdot\vec{p}_\mathrm{T}^\ell\right)^\frac{1}{2}$.
Since it is predominantly through hadronic $\tau$ decays that $W$ bosons and \ttbar{} pairs contribute
to the signal regions, the corresponding control regions are created by recasting the muon
or electron as a jet.
If that lepton has sufficient \pt{} (without any additional calibration)
it may contribute to the jet multiplicity count
(denoted \nfiftyCR or \neightyCR), as well as to \HT{}
and hence to \metsig{}.
In order to yield sufficient numbers of events in these
CRs, the requirement
on the jet multiplicity in each CR is one fewer
than that in the corresponding SR, and
a somewhat less stringent requirement is made on \metsig{} compared to the SRs.

For each SR (regardless of its own requirement on \nbjet)
there are two CRs, which require either exactly zero or at least
one \bjet.
These help constrain the combination of \ttbar{} and $W$+jets
backgrounds, with the \ttbar{} background being enhanced in the CR that requires a \bjet.
Figure~\ref{fig:CR:njet} shows the resulting \nfiftyCR{} jet multiplicity distributions in these control regions.

\begin{table}
\centering\renewcommand\arraystretch{1.3}
\begin{tabular}{|l||c|c|c|c|}
\hline
\textbf{SR name} & \multicolumn{2}{c|}{\hfill \boldSR{$n$j50} or \boldSR{$n$j50-1b} or \boldSR{$n$j50-2b} \hfill}
                 & \multicolumn{2}{c|}{\hfill \boldSR{$n$j80} or \boldSR{$n$j80-1b} or \boldSR{$n$j80-2b} \hfill} \\
\hline
\textbf{CR name} & \boldSR{CR$(n-1)$j50-0b} & \boldSR{CR$(n-1)$j50-1b} & \boldSR{CR$(n-1)$j80-0b} & \boldSR{CR$(n-1)$j80-1b} \\
\hline\hline
$\pt^{\ell}$ ($\ell\in\{e\,\mu\}$) & \multicolumn{4}{c|}{$>20\GeV$}\\
\hline
\mt        & \multicolumn{4}{c|}{$<120\GeV$}\\
\hline
\metsig    & \multicolumn{4}{c|}{$>3\rootgev$} \\
\hline
\nfiftyCR  & \multicolumn{2}{c|}{$\geq \nfifty-1$} & \multicolumn{2}{c|}{---} \\
\hline
\neightyCR & \multicolumn{2}{c|}{---} & \multicolumn{2}{c|}{$\geq \neighty-1$} \\
\hline
\nbjet     & 0 & $\geq 1$ & 0 & $\geq 1$\\
\hline
\end{tabular}
\caption{\label{tab:CRdefs}
Leptonic control region definitions for each of the signal regions.
In the names, the symbols $n$ and $n-1$ refer to the
corresponding jet multiplicity requirements. For example the three signal regions
\SR{9j50}, \SR{9j50-1b} and \SR{9j50-2b} are each independently controlled by both the
\SR{CR8j50-0b} and \SR{CR8j50-1b} control regions.
}
\end{table}

\label{sec:fit}
For each signal region, a simultaneous fit is performed
to the number of events found in the corresponding two CRs,
using the HistFitter package~\cite{Baak:2014wma}. For the purpose of exclusion,
the simultaneous fit also includes data in the SR.
In the fit the normalisations of the \ttbar{} and $W$+jets background contributions
are allowed to float, while the other \leptonic{} backgrounds, which are
generally subdominant,
are determined directly from their yields using the corresponding theoretical cross-sections.
The event yields in each CR and SR are assumed to be Poisson distributed.
The systematic uncertainties are treated as Gaussian-distributed nuisance parameters,
and are assumed to be correlated within each fit.
The \multijet{} background yield in the SR is determined
separately from the data using the methods described in Section~\ref{sec:bg:multijet}.

The normalisations for the \ttbar{} and $W$+jets backgrounds
are generally found to be consistent with their corresponding theoretical predictions
when uncertainties are considered.
Systematic uncertainties are larger than statistical uncertainties for the regions with looser selection criteria,
with the situation reversed for those with tighter selection criteria.
The systematic uncertainties with the largest impact include theoretical uncertainties on the \ttbar{} background,
the impact of limited numbers of events in the control regions, the
closure of the multijet background estimation method
 and the jet energy scale.
The overall post-fit values range from 14\% to 42\% with the theoretical uncertainties
on the \ttbar{} backgrounds typically being the most significant contribution.

\oursection{Results}\label{sec:results}

\begin{table}
\centering
\begin{tabular}{@{}>{\small\bfseries}lr@{${}\pm{}$}lr@{${}\pm{}$}lr@{${}\pm{}$}lc@{}}
\toprule

\multirow{2}{*}{Signal region} & \multicolumn{6}{c}{Fitted background}
      & \multirow{2}{*}{Obs events} \\ \cline{2-7}
                             & \multicolumn{2}{c}{Multijet} &
\multicolumn{2}{c}{Leptonic} & \multicolumn{2}{c}{Total} &\\

\midrule

\SR{8j50}     & $109.3$ & $6.9$  & $80$   & $25$   & $189$  & $26$   & $157$ \\
\SR{8j50-1b}  & $76.7$  & $2.7$  & $62$   & $21$   & $138$  & $21$   & $97$  \\
\SR{8j50-2b}  & $33.8$  & $2.1$  & $33$   & $13$   & $67$   & $13$   & $39$  \\
\SR{9j50}     & $16.8$  & $1.3$  & $12.8$ & $5.4$  & $29.6$ & $5.6$  & $29$  \\
\SR{9j50-1b}  & $13.5$  & $2.0$  & $10.2$ & $4.9$  & $23.8$ & $5.3$  & $21$  \\
\SR{9j50-2b}  & $6.4$   & $1.6$  & $5.8$  & $3.3$  & $12.1$ & $3.6$  & $9$   \\
\SR{10j50}    & $2.61$  & $0.61$ & $1.99$ & $0.62$ & $4.60$ & $0.87$ & $6$   \\
\SR{10j50-1b} & $2.42$  & $0.62$ & $1.44$ & $0.49$ & $3.86$ & $0.79$ & $3$   \\
\SR{10j50-2b} & $1.40$  & $0.87$ & $0.83$ & $0.37$ & $2.23$ & $0.94$ & $1$   \\
\addlinespace
\SR{7j80}     & $40.0$  & $5.3$  & $30$   & $13$   & $70$   & $14$   & $70$  \\
\SR{7j80-1b}  & $29.1$  & $3.4$  & $20.8$ & $10$   & $50$   & $11$   & $42$  \\
\SR{7j80-2b}  & $11.5$  & $1.6$  & $11.0$ & $5.0$  & $22.5$ & $5.2$  & $19$  \\
\SR{8j80}     & $4.5$   & $1.9$  & $4.9$  & $2.2$  & $9.3$  & $2.9$  & $8$   \\
\SR{8j80-1b}  & $3.9$   & $1.5$  & $3.8$  & $2.1$  & $7.6$  & $2.6$  & $4$   \\
\SR{8j80-2b}  & $1.72$  & $0.93$ & $2.3$  & $1.1$  & $4.1$  & $1.5$  & $2$   \\

\bottomrule
\end{tabular}
\caption{For each signal region, the expected SM background (and separately the multijet and leptonic contributions)
and the observed number of data events. The SM background normalisations are obtained from fits to the data in control regions,
as described in Sections~\ref{sec:bg} and \ref{sec:systematics}. The signal regions are as defined in Table~\ref{tab:SRdefs}. }
\label{tab:fitResults}
\end{table}

\begin{figure}\begin{center}
\centering
\subfigure[\nfifty]{\includegraphics[width=0.75\textwidth]{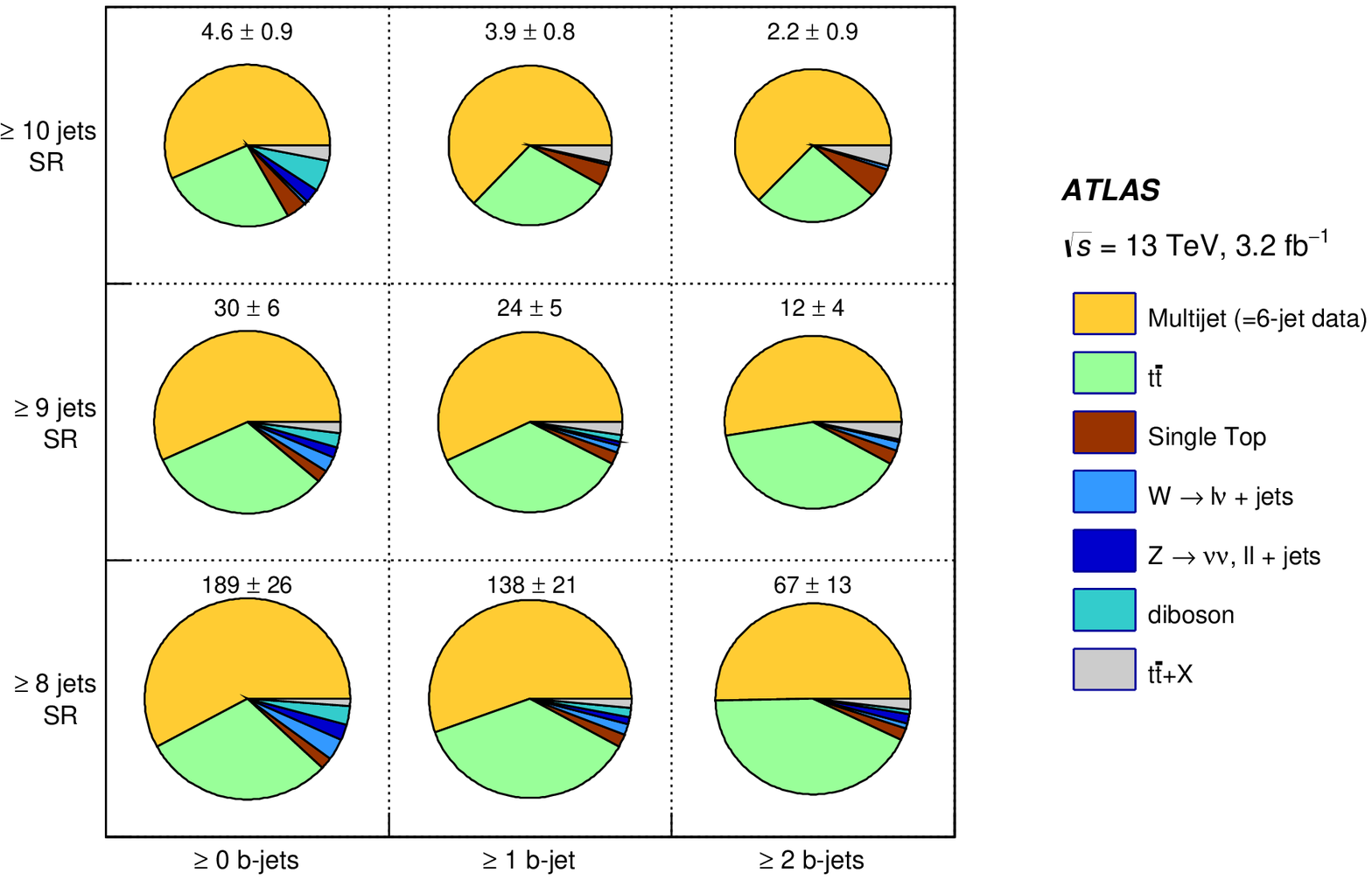}} \\
\subfigure[\neighty]{\includegraphics[width=0.75\textwidth]{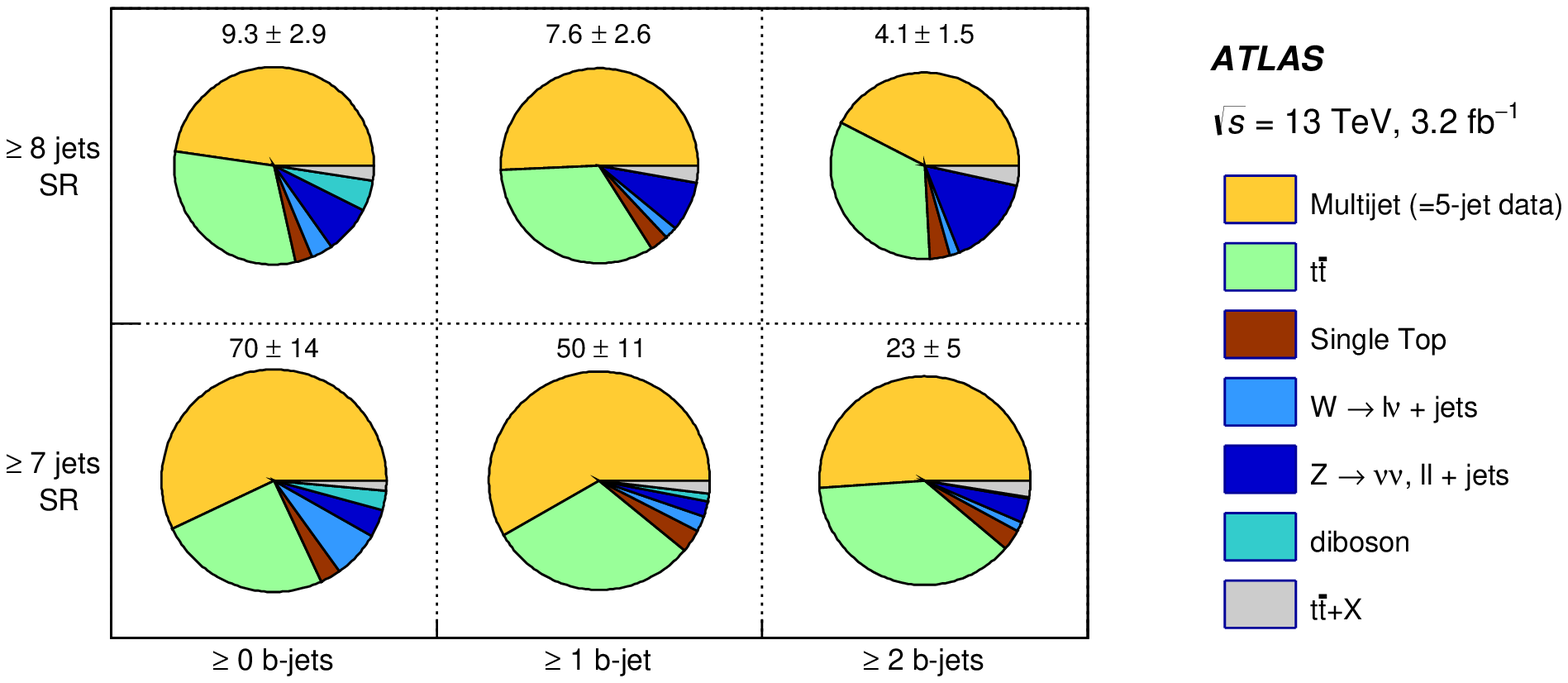}}
\caption{\label{fig:pies:SR}
Post-fit signal region compositions.
The area of each pie chart is scaled to log$_{10}$ of the total expected yield (as printed above each one).
}
\end{center}\end{figure}

Figure~\ref{fig:metsig_final} shows the post-fit \metsig distributions in the most sensitive signal regions (see below), while
Figure~\ref{fig:pies:SR} shows the background composition in all fifteen SRs.
The background is split between multijet
and \leptonic{} processes, with the latter being 60--90\% \ttbar{}.

The yields in each of the 15 signal regions are reported in
Table~\ref{tab:fitResults}.
No significant excess is observed above the SM expectations in any SR,
and most have confidence levels for the background-only hypothesis larger than 10\%,
as shown in Table~\ref{tab:upperLimits}.
The table also shows the model-independent limits -- 95\% confidence level (CL) limits
on the maximum contribution of new physics processes to the event yields in the various SRs,
assuming  zero signal contamination in control regions.

\begin{table}
\begin{center}
\setlength{\tabcolsep}{0.0pc}
\begin{tabular*}{\textwidth}{@{\extracolsep{\fill}}lccccc}
\noalign{\smallskip}\hline\noalign{\smallskip}
{\bf Signal region}  & $\langle\epsilon{\rm \sigma}\rangle_{\rm obs}^{95}$[fb]  &  $S_{\rm obs}^{95}$  & $S_{\rm exp}^{95}$ & $CL_{B}$ & $p(s=0)$  \\
\noalign{\smallskip}\hline\noalign{\smallskip}

\SR{8j50}    & $11$ &  $36$  & ${49}^{+19}_{-13}$ &  $0.14$  &    $0.50$  \\
 \SR{8j50-1b}    & $6.8$ &  $22$  & ${37}^{+13}_{-10}$ &  $0.04$  &    $0.50$  \\
 \SR{8j50-2b}    & $3.8$ &  $12$  & ${22}^{+8}_{-6}$ &  $0.03$  &    $0.50$  \\
 \SR{9j50}    & $5.8$ &  $19$  & ${19}^{+4}_{-5}$ &  $0.49$  &    $0.50$  \\
 \SR{9j50-1b}    & $5$ &  $16$  & ${17}^{+2}_{-6}$ &  $0.38$  &    $0.50$  \\
 \SR{9j50-2b}    & $2.6$ &  $8$  & ${10}^{+3}_{-2}$ &  $0.31$  &    $0.50$  \\
 \SR{10j50}    & $2.5$ &  $8$  & ${6}^{+3}_{-1}$ &  $0.74$  &    $0.26$  \\
 \SR{10j50-1b}    & $1.6$ &  $5$  & ${6}^{+2}_{-1}$ &  $0.37$  &    $0.50$  \\
 \SR{10j50-2b}    & $1.1$ &  $4$  & ${4}^{+2}_{-1}$ &  $0.27$  &    $0.50$  \\
 \addlinespace
\SR{7j80}    & $10$ &  $32$  & ${32}^{+11}_{-9}$ &  $0.51$  &    $0.50$  \\
 \SR{7j80-1b}    & $6.2$ &  $20$  & ${24}^{+6}_{-5}$ &  $0.29$  &    $0.50$  \\
 \SR{7j80-2b}    & $4.2$ &  $14$  & ${14}^{+6}_{-2}$ &  $0.33$  &    $0.50$  \\
 \SR{8j80}    & $3.2$ &  $10$  & ${11}^{+2}_{-4}$ &  $0.41$  &    $0.50$  \\
 \SR{8j80-1b}    & $1.7$ &  $5$  & ${7}^{+3}_{-2}$ &  $0.20$  &    $0.50$  \\
 \SR{8j80-2b}    & $1.4$ &  $4$  & ${5}^{+2}_{-1}$ &  $0.24$  &    $0.50$  \\

\noalign{\smallskip}\hline\noalign{\smallskip}
\end{tabular*}
\end{center}
\caption{
        The results of a fit to the control and signal region data, assuming no
        signal contamination in the control regions.
        Left to right: 95\% CL upper limits on the visible cross-section
        ($\langle\epsilon\sigma\rangle_{\rm obs}^{95}$) and on the number of
        signal events ($S_{\rm obs}^{95}$ ).
        Convergence and stability tests of the fits suggest uncertainties of order 5\% on $S_{\rm obs}^{95}$ resulting from these effects.
        The third column
        ($S_{\rm exp}^{95}$) shows the 95\% CL upper limit on the number of
        signal events, given the expected number (and $\pm 1\sigma$
            excursions on the expectation) of background events.
        The last two columns
        indicate the $CL_B$ value, i.e. the confidence level observed for
        the background-only hypothesis, and the discovery $p$-value ($p(s = 0)$).
        The test is one-sided, so the $p$-value is 0.50 when the observed number of events is smaller than the prediction.
        Yields are not statistically independent, since there are correlated systematic uncertainties and since signal regions overlap.
        \label{tab:upperLimits}}
\end{table}

\begin{figure}
\centering
\subfigure[pMSSM slice]
{\includegraphics[width=0.65\textwidth]{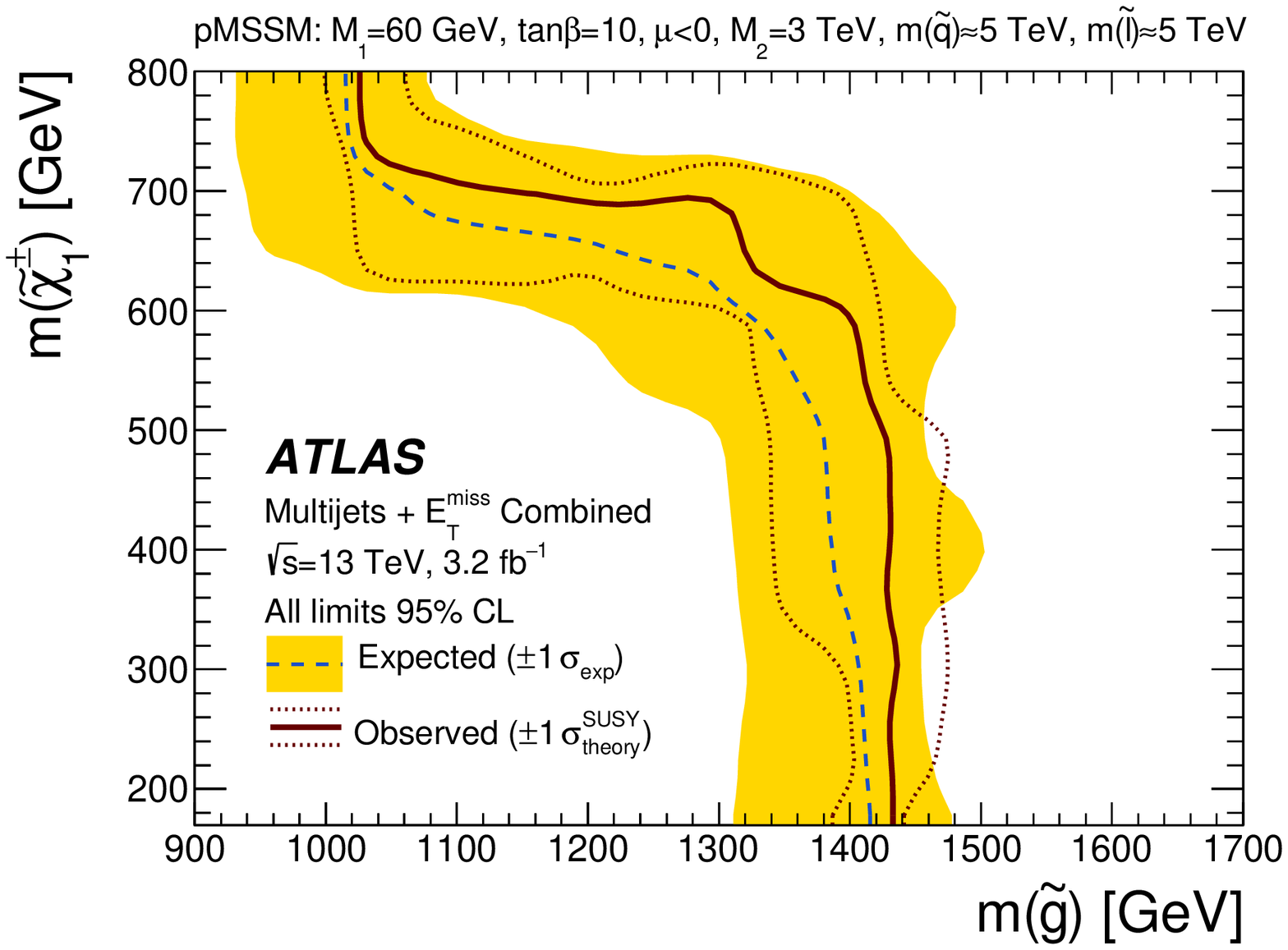}}
\subfigure[Simplified cascade decay (`2-step') model]{\includegraphics[width=0.65\textwidth]{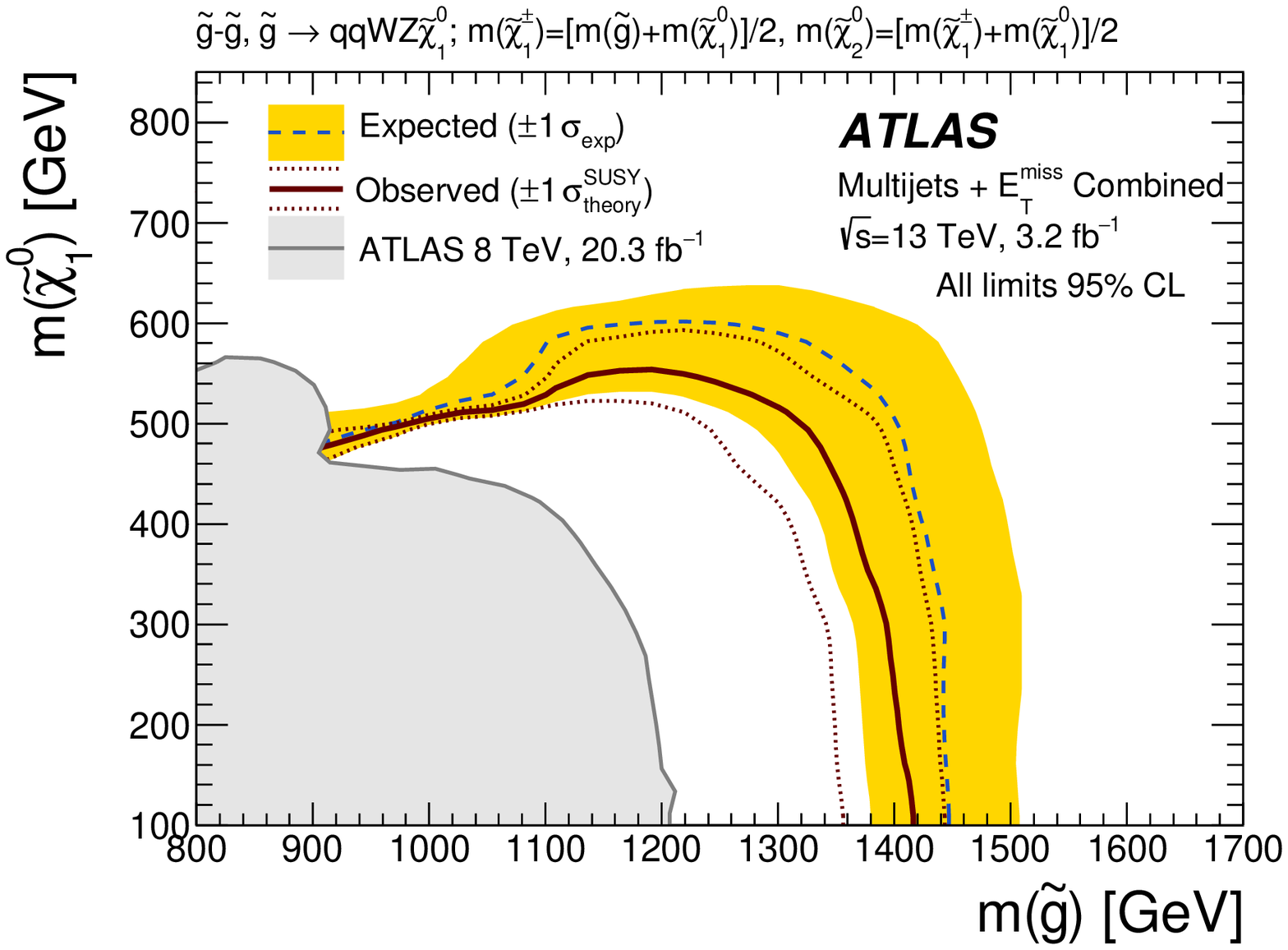}}
\caption{
95\% CL exclusion curve for the two supersymmetric models described in the text.
The solid red and dashed blue curves show the 95\% CL observed and expected limits, respectively,
including all uncertainties except the theoretical signal cross-section uncertainty
(PDF and scale).
The dotted red lines bracketing the observed limit represent the result produced when moving
the signal cross-section by $\pm1\sigma$ (as defined by the PDF and scale uncertainties).
The shaded yellow band around the expected limit shows the
$\pm 1\sigma$ variation of the expected limit.
The shaded grey area shows the observed
exclusion from the combination of ATLAS $\sqrt{s}=8\TeV$ analyses performed in Ref. \cite{Aad:2015iea} (Figure~25 therein).
Excluded regions are below and to the left of the relevant lines.
}
\label{fig:contour:both}
\end{figure}

The results are interpreted in the
context of the two supersymmetric models described in Section~\ref{sec:signals}.
The limit for each signal region is obtained by comparing the observed event count
with that expected from Standard Model background plus SUSY signal processes, with
their contamination of the leptonic control regions, typically below 10\% for points close to the exclusion contour,
being accounted for.
All uncertainties on the Standard Model expectation are considered, including those which are correlated
between signal and background (for instance jet energy scale uncertainties) and all,
except theoretical cross-section uncertainties (PDF and scale), on the signal expectation.
The resulting exclusion regions, shown in Figure~\ref{fig:contour:both},
are obtained using the \CLs prescription~\cite{clsread}.
For each signal model point, the signal region with the best expected limit is used.
Signal regions defined by \nfifty{} and those defined by \neighty{}
both contribute to the best expected limit.
The most sensitive signal regions are found to be those with no requirement on $\nbjet{}$
for the simplified model decay.
For the pMSSM slice, which has large branching ratios for gluinos to
third-generation quarks, the best signal regions are those requiring either one or two \bjets.
In both cases, gluino masses up to $1400\GeV$ are excluded at 95\% confidence level,
significantly extending previous limits for the simplified model decay.

\FloatBarrier
\oursection{Conclusions}\label{sec:conclusions}

A search is presented for new phenomena with large jet multiplicities (from $\ge 7$ to $\ge 10$)
and missing transverse momentum. The search used 3.2\,fb$^{-1}$ of $\sqrt{s}=13\TeV$ $pp$ collision data
collected by the ATLAS experiment at the Large Hadron Collider.
The increase in the LHC centre-of-mass energy provided increased sensitivity to
higher-mass sparticles compared with previous searches.
Further sensitivity was gained by considering separately final states with
$\ge 0$, $\ge 1$ and $\ge 2$ \btagged{} jets.
The Standard Model predictions are found to be consistent with the data.
The results are interpreted in the context of a simplified
supersymmetry model, and a slice of the pMSSM, each of
which predict cascade decays of supersymmetric particles and hence large jet multiplicities.
The data exclude gluino masses up to $1400\GeV$ at the 95\% CL in these models,
significantly extending previous bounds.
Model-independent limits were presented which allow reinterpretation of the results
to cases of other models which also predict decays into multijet final states
in association with invisible particles.

\begin{small}
\section*{Acknowledgements}

We thank CERN for the very successful operation of the LHC, as well as the
support staff from our institutions without whom ATLAS could not be
operated efficiently.

We acknowledge the support of ANPCyT, Argentina; YerPhI, Armenia; ARC, Australia; BMWFW and FWF, Austria; ANAS, Azerbaijan; SSTC, Belarus; CNPq and FAPESP, Brazil; NSERC, NRC and CFI, Canada; CERN; CONICYT, Chile; CAS, MOST and NSFC, China; COLCIENCIAS, Colombia; MSMT CR, MPO CR and VSC CR, Czech Republic; DNRF and DNSRC, Denmark; IN2P3-CNRS, CEA-DSM/IRFU, France; GNSF, Georgia; BMBF, HGF, and MPG, Germany; GSRT, Greece; RGC, Hong Kong SAR, China; ISF, I-CORE and Benoziyo Center, Israel; INFN, Italy; MEXT and JSPS, Japan; CNRST, Morocco; FOM and NWO, Netherlands; RCN, Norway; MNiSW and NCN, Poland; FCT, Portugal; MNE/IFA, Romania; MES of Russia and NRC KI, Russian Federation; JINR; MESTD, Serbia; MSSR, Slovakia; ARRS and MIZ\v{S}, Slovenia; DST/NRF, South Africa; MINECO, Spain; SRC and Wallenberg Foundation, Sweden; SERI, SNSF and Cantons of Bern and Geneva, Switzerland; MOST, Taiwan; TAEK, Turkey; STFC, United Kingdom; DOE and NSF, United States of America. In addition, individual groups and members have received support from BCKDF, the Canada Council, CANARIE, CRC, Compute Canada, FQRNT, and the Ontario Innovation Trust, Canada; EPLANET, ERC, FP7, Horizon 2020 and Marie Sk{\l}odowska-Curie Actions, European Union; Investissements d'Avenir Labex and Idex, ANR, R{\'e}gion Auvergne and Fondation Partager le Savoir, France; DFG and AvH Foundation, Germany; Herakleitos, Thales and Aristeia programmes co-financed by EU-ESF and the Greek NSRF; BSF, GIF and Minerva, Israel; BRF, Norway; the Royal Society and Leverhulme Trust, United Kingdom.

The crucial computing support from all WLCG partners is acknowledged
gratefully, in particular from CERN and the ATLAS Tier-1 facilities at
TRIUMF (Canada), NDGF (Denmark, Norway, Sweden), CC-IN2P3 (France),
KIT/GridKA (Germany), INFN-CNAF (Italy), NL-T1 (Netherlands), PIC (Spain),
ASGC (Taiwan), RAL (UK) and BNL (USA) and in the Tier-2 facilities
worldwide.

\end{small}

\FloatBarrier

\printbibliography

\newpage
\begin{flushleft}
{\Large The ATLAS Collaboration}

\bigskip

G.~Aad$^\textrm{\scriptsize 86}$,
B.~Abbott$^\textrm{\scriptsize 113}$,
J.~Abdallah$^\textrm{\scriptsize 151}$,
O.~Abdinov$^\textrm{\scriptsize 11}$,
B.~Abeloos$^\textrm{\scriptsize 117}$,
R.~Aben$^\textrm{\scriptsize 107}$,
M.~Abolins$^\textrm{\scriptsize 91}$,
O.S.~AbouZeid$^\textrm{\scriptsize 137}$,
H.~Abramowicz$^\textrm{\scriptsize 153}$,
H.~Abreu$^\textrm{\scriptsize 152}$
% ...
\bigskip
\\
% ...
$^{11}$ Institute of Physics, Azerbaijan Academy of Sciences, Baku, Azerbaijan\\
% ...
$^{86}$ CPPM, Aix-Marseille Universit{\'e} and CNRS/IN2P3, Marseille, France\\
% ...
$^{91}$ Department of Physics and Astronomy, Michigan State University, East Lansing MI, United States of America\\
% ...
$^{107}$ Nikhef National Institute for Subatomic Physics and University of Amsterdam, Amsterdam, Netherlands\\
% ...
$^{113}$ Homer L. Dodge Department of Physics and Astronomy, University of Oklahoma, Norman OK, United States of America\\
% ...
$^{117}$ LAL, Univ. Paris-Sud, CNRS/IN2P3, Universit{\'e} Paris-Saclay, Orsay, France\\
% ...
$^{137}$ Santa Cruz Institute for Particle Physics, University of California Santa Cruz, Santa Cruz CA, United States of America\\
% ...
$^{151}$ Institute of Physics, Academia Sinica, Taipei, Taiwan\\
$^{152}$ Department of Physics, Technion: Israel Institute of Technology, Haifa, Israel\\
$^{153}$ Raymond and Beverly Sackler School of Physics and Astronomy, Tel Aviv University, Tel Aviv, Israel\\
% ...
\end{flushleft}


\clearpage

\end{document}